\documentclass[aps,twocolumn,raggedbottom,prl,showpacs,nobalancelastpage,amssymb,superscriptaddress]{revtex4} 
\usepackage[english]{babel}
\usepackage{graphicx}
\usepackage{amssymb}
\usepackage{amsmath}
\usepackage{amscd}
\usepackage{eucal}
\usepackage{color}
\usepackage{bm}
\usepackage{upgreek}

\usepackage{epstopdf}

\def\beq{\begin{equation}}
\def\eeq{\end{equation}}
\def\beqa{\begin{eqnarray}}
\def\eeqa{\end{eqnarray}}
\def\bsub{\begin{subequations}}
\def\esub{\end{subequations}}

\newcommand{\ci}{\mathrm{i}}
\def\mathvecfont#1{\textbf{\em #1}}
\newcommand{\vek}[1]{\mathvecfont{#1}}
\newcommand{\op}[1]{\hat{#1}}

\begin{document}

\title{Weak localization in mesoscopic hole transport: \\
          Berry phases and classical correlations }

\author{Viktor Krueckl}
\affiliation{Institut f\"ur Theoretische Physik, Universit\"at Regensburg, D-93040 Regensburg, Germany}
\author{Michael Wimmer}
\affiliation{Instituut-Lorentz, Universiteit Leiden, P.O. Box 9506, 2300 RA Leiden, The Netherlands}
\author{\.{I}nan\c{c} Adagideli}
\affiliation{Faculty of Engineering and Natural Sciences, Sabanci University, Istanbul 34956, Turkey}
\author{Jack Kuipers}
\affiliation{Institut f\"ur Theoretische Physik, Universit\"at Regensburg, D-93040 Regensburg, Germany}
\author{Klaus Richter}
\affiliation{Institut f\"ur Theoretische Physik, Universit\"at Regensburg, D-93040 Regensburg, Germany}

\date{\today}

\begin{abstract}
We consider phase-coherent transport through ballistic and diffusive two-dimensional
hole systems based on the Kohn-Luttinger Hamiltonian. We show that
intrinsic heavy-hole light-hole coupling gives rise to clear-cut signatures of an
associated Berry phase in the weak localization which renders the magneto-conductance
profile distinctly different from electron transport.
Non-universal classical correlations determine the strength of these Berry phase effects
and the effective symmetry class,
leading even to antilocalization-type features for circular quantum dots and Aharonov-Bohm
rings in the absence of additional spin-orbit interaction. Our semiclassical predictions are quantitatively confirmed by
numerical transport calculations.
\end{abstract}



\maketitle

As a genuine wave phenomenon, coherent backscattering, denoting enhanced backreflection
of waves in complex media due to constructive interference of time-reversed paths,
has been encountered in numerous systems. Its occurrence ranges from the
observation of the infrared intensity reflected from Saturn's rings \cite{Hapke1993} to light
scattering in random media \cite{Albada-Wolf}, from enhanced backscattering of
seismic \cite{Larose2004} and acoustic \cite{Bayer1993} to atomic matter waves \cite{Hartung2008}.
In condensed matter, weak localization (WL) \cite{Abrahams1979,Altshuler1980},
closely related to coherent backscattering, has been widely used as a diagnostic tool for
probing phase coherence in conductors at low temperatures. Based on time-reversal symmetry (TRS), WL
manifests itself as a characteristic dip in the average magneto conductivity
at zero magnetic field $B$.
The opposite phenomenon, a peak at $B\!=\!0$, is usual\-ly interpreted as weak antilocalization (WAL)
due to spin-orbit interaction (SOI) \cite{Hikami1980}.

In this Letter we show that the average magneto conductance of mesoscopic systems
built from two-dimensional hole gases (2DHG) distinctly deviates from the corresponding WL
transmission dip profiles of their n-doped counterparts.
In particular, ballistic hole conductors such as circular quantum dots and
Aharonov-Bohm (AB) rings, can exhibit a conductance peak at $B\!=\!0$, even in the absence of SOI
 \cite{winkler:book} due to structure (SIA) or bulk inversion (BIA) asymmetry.
We trace this back to effective TRS breaking of hole systems at $B\!=\!0$.

Recently, various magnetotransport measurements on such high-mobility 2DHG
have been performed for GaAs bulk samples \cite{McPhail2004},
quasi-ballistic cavities \cite{Faniel2007} and AB rings \cite{Yau2002,Grbic2007}.
However, we are not aware of corresponding theoretical approaches for ballistic
2DHG nanoconductors (except for 1d models \cite{Zulicke2010}),
despite the huge number of theory works on ballistic electron transport \cite{Ferry1997,Beenakker1997}.
Here we treat 2DHG-based ballistic and diffusive mesoscopic
structures on the level of the 4-band Kohn-Luttinger Hamiltonian \cite{luttinger1955}.
By devising a semiclassical approach for ballistic, coupled heavy-hole (HH) light-hole (LH)
dynamics we
can associate the anomalous WL features directly with Berry phases \cite{Berry84}
in the Kohn-Luttinger model \cite{Chang96,haldane04,Chang2008} (that have proven
relevant e.g.\ for the spin Hall effect \cite{Murakami2003}).
We show that the strength of the related effective 'Berry field',
giving rise to effective TRS breaking and a splitting of the WL dip,
is determined by a classical correlation between enclosed areas and reflection angles
of interfering hole trajectories relevant for WL.
This system-dependent geometrical correlation is not amenable to existing
random matrix approaches for chaotic conductors \cite{Beenakker1997}.
We confirm our semiclassical results by numerical quantum
transport calculations and further discuss the additional effect of SOI.

{\em Hamiltonian and band structure.--}
To describe the 2DHG we represent the Kohn-Luttinger Hamiltonian \cite{luttinger1955}
for the two uppermost valence bands of a semiconductor
in terms of an eigenmode expansion for an infinite square well of width $a$
modelling the vertical confinement.
Employing L{\"o}wdin partitioning \cite{loewdin1951} we construct an effective
Hamiltonian based on the relevant, lowest subband in $z$-direction
\cite{comment-subband}.
The resulting $4\!\times \! 4$-Luttinger Hamiltonian for a quasi 2DHG then describes coupled
HH and LH states with spin projection $\pm 3/2$, and $\pm 1/2$, respectively.
Without SOI due to SIA or BIA, the 2DHG Hamiltonian splits into decoupled blocks:
%
\begin{equation}
\op{\mathcal H}_\mathrm{2D} =
  \begin{pmatrix}
        \op P & \op T &  &  \\
        \op T^\dagger & \op Q &  &  \\
        &  & \op Q &  \op T\\
        &  & \op T^\dagger & \op P
  \end{pmatrix} =
  \begin{pmatrix}
        \op{\mathcal H}_\mathrm{U} &  \\
        & \op{\mathcal H}_\mathrm{L}\\
  \end{pmatrix}
  \text{\hspace*{0.3cm}} , \,
  \begin{matrix}
        \mathrm{HH} \Uparrow \\
        \mathrm{LH} \downarrow \\
        \mathrm{LH} \uparrow \\
        \mathrm{HH} \Downarrow
  \end{matrix}
  \label{2Dapprox}
\end{equation}
%
with the upper and lower blocks
composed of \cite{broido1985}
%
%
\begin{subequations}
\label{lut:firstorder}
\begin{eqnarray}
 \op P &=& -\frac{\hbar^2}{2m_0} \left[  (\gamma_1+\gamma_2)\op k_\Vert^2
         + (\gamma_1-2\gamma_2) \langle \op k_z^2 \rangle \right]
\text{,}\\
 \op Q &=& -\frac{\hbar^2}{2m_0} \left[  (\gamma_1-\gamma_2) \op k_\Vert^2
         + (\gamma_1+2\gamma_2) \langle \op k_z^2 \rangle \right]
\text{,}\\
 \op T &=& -\sqrt{3} \frac{\hbar^2}{2m_0}
   \left[
      \gamma_2 (\op k_x^2- \op k_y^2) + 2 \mathrm{i} \gamma_3 \op k_x \op k_y
   \right ]
\text{.}
\end{eqnarray}
\end{subequations}
%
Here,  $\op {\bf k} = (\op k_x, \op k_y, \op k_z)$ is the wave vector
with projection  $\op k_\Vert$  onto the $xy$-plane of the 2DHG and
$\langle \op k_z^2 \rangle \!=\! (\pi/a)^2$ is  the expectation value of
$k_z$ for the lowest subband.
Below we use the axial approximation,
$\bar \gamma\!=\! \gamma_2 \!=\! \gamma_3$, for the parameters in
$\op T$ that couple HH and LH states.

Due to the 2D confinement the HH-LH bulk degeneracy
is lifted which will play an important role for the WL analysis below.
To this end we will calculate the two-terminal Landauer conductance
%
\begin{equation}
G  =\frac{e^2}{h} T
 =\frac{e^2}{h}\sum^N_{n,m}\sum_{\sigma, \sigma'}
\vert t_{m, \sigma'; n, \sigma} \vert^2
\label{Landauer}
\end{equation}
%
with the transmission amplitudes $t_{m, \sigma'; n, \sigma}$
given by the Fisher-Lee relations~\cite{fischer1981}.
The indices $m$ and $n$ label $N$ transverse modes in the leads, and
 $\sigma \in \{ \mathrm{U}, \mathrm{L} \}$
with $\mathrm{U} \in \{ {\mathrm{HH} \Uparrow, \mathrm{LH} \downarrow} \}$
and $\mathrm{L} \in \{ \mathrm{HH} \Downarrow, \mathrm{LH} \uparrow \}$
denotes the HH and LH modes.
The Hamiltonian~(\ref{2Dapprox}) with blocks obeying
$ \op{\mathcal H}_\mathrm{U}(B) \!=\! \op{\mathcal H}^\dagger_\mathrm{L}(-B) $
(neglecting Zeeman spin splitting) allows us to
separately define related total transmissions, $ T_\mathrm{U},  T_\mathrm{L}$,
with
$T \!=\!  T_\mathrm{U} \!+\! T_\mathrm{L}$
fulfilling $T_\mathrm{U}(B) \!=\! T_\mathrm{L}(-B)$.

Depending on the position of the Fermi level $E_{\rm F}$ we distinguish the case where
HH and LH states are both occupied (considered at the end of this Letter)
from the case where $E_{\rm F}$ is close to the band gap such
that only HH states contribute to transport. We first study the latter case with focus
on effects from the HH-LH coupling.

{\em HH-LH coupling and Berry phase.-}
For ballistic mesoscopic systems of linear size $L$ in the regime $kL\! \gg\! 1$ we will
generalize the semiclassical approaches \cite{Baranger1993,Richter2002} to the Landauer conductance
from electron systems with a parabolic dispersion
to the p-doped case with more complex band topology. The HH-LH coupling enters into
the semiclassical formalism as an additional phase that is accumulated during each
reflection of a HH wave packet at a smooth boundary potential (the hard wall case is
considered below).  Such a reflection can be described as an adiabatic transition
in momentum space leading to a geometric phase acquired along a given path
\cite{Chang96,haldane04}:
\begin{equation}
\Gamma_\sigma\!=\!\int\! \mathcal{A}_\sigma(\vek{k}) d\vek{k} \quad ;  \quad
 \mathcal{A}_\sigma(\vek{k}) \! = \!
    -\ci\langle \psi_\sigma(\vek k) \vert \nabla_k \psi_\sigma(\vek k) \rangle  \, .
\label{berry-general}
\end{equation}
Using for $ \psi_\sigma(\vek k)$ the free solutions of Hamiltonian~(\ref{2Dapprox})
we find after diagonalization for the vector potential
\begin{equation}
\mathcal A_{\mathrm{HH} \Uparrow }(\vek k) =
-\mathcal A_{\mathrm{HH} \Downarrow }(\vek k) =
3 \frac{\xi^{\rm Berry} (k) }{k^2}
\begin{pmatrix}
k_y \\ -k_x
\end{pmatrix}
\label{berryA}
\end{equation}
and
$\mathcal A_{\mathrm{LH} \downarrow }(\vek k) =
\mathcal -A_{\mathrm{LH} \uparrow }(\vek k) =
-[(3\xi+2)/3\xi] \mathcal A_{\mathrm{HH} \Uparrow }(\vek k)$
with
$ \xi^{\rm Berry} (k) \!  \simeq \! -\tfrac{1}{8} ( \tfrac{k a}{\pi} )^4 $,
to leading order in $k a/\pi$.
The Berry phase for a single reflection at a smooth boundary is then
\begin{equation}
\label{berryphase}
\Gamma_{\mathrm{HH} \Uparrow }^{\rm Berry}   (\varphi) = -\Gamma_{\mathrm{HH} \Downarrow }^{\rm Berry} ( \varphi)
   =  \xi^{\rm Berry} \sin \varphi (2 - \cos \varphi )  \, ,
\end{equation}
where $ \varphi$ denotes the change in momentum direction.

For a specular reflection at a hard-wall (hw) confinement  a corresponding
phase shift
is obtained by requiring that
the propagating HH and the evanescent LH part
of the reflected wave both must vanish at the boundary:
\begin{eqnarray}
\Gamma_{\mathrm{HH} \Uparrow }^{\rm hw}  (\varphi) & \!=\! &
 \frac{1}{\ci}\ln  \frac{2 - \xi^{\rm hw} e^{-2 \ci \, \varphi}}{\vert 2 - \xi^{\rm hw}
e^{-2 \ci \, \varphi} \vert}
\stackrel{\xi^{\rm hw} \ll 1}{\simeq}   \xi^{\rm hw} \sin  2\varphi  \, ,
  \label{reflectionphase}   \\
\mbox{ with } \quad  & &
 \xi^{\rm hw} (k)   \simeq
 -\frac{\gamma_1 + \bar \gamma}{4\bar \gamma} \left( \frac{k a}{\pi} \right)^2  \, .
\label{approxfraction}
\end{eqnarray}

{\em Average magneto conductance.-} A semiclassical approach proves convenient
to incorporate these additional (Berry) phases
into a theory of WL.
For a (chaotic) ballistic quantum dot the known semiclassical amplitude
\cite{Baranger1993} for electron transmission from channel $n$ to $m$ is generalized
to
$ t_{m,\mathrm{HH}\Uparrow;n,\mathrm{HH}\Uparrow} \!\simeq\!
 \sum_\gamma C_\gamma K_\gamma \exp(\tfrac{\ci}{\hbar} \mathcal S_\gamma)$,
in terms of a sum over lead-connecting classical paths $\gamma$ with
classical action $\mathcal S_\gamma $,
weight $C_\gamma$ (including the Maslov index) and an additional factor
$K_\gamma\!=\! \exp[\ci \sum_{j=1}^{n_b} \Gamma_{\mathrm{HH}\Uparrow} (\varphi_j)]$
accounting for the accumulated phases (\ref{berryphase}) or (\ref{reflectionphase})
after $n_b$ successive reflections.
In view of Eq.~(\ref{Landauer}) the total semiclassical transmission probability
 for HH$\Uparrow$ states reads
\begin{equation}
  T_{\rm U} \simeq  \sum_{n,m} \sum_{\gamma \gamma'}
 K_\gamma K^*_{\gamma'}  C_{\gamma} C^*_{\gamma'}
       \mathrm{e}^{\tfrac{\ci}{\hbar} (\mathcal S_\gamma - \mathcal S_{\gamma'})}
\mathrm{.}
\label{semiclT}
\end{equation}
The  diagonal contribution, $\gamma \!=\! \gamma'$, correctly yields the
classical transmission since $ K_{\gamma} K^{*}_{\gamma} \!=\! 1$.
WL contributions arise (after averaging)
from off-diagonal pairs of long, classically correlated paths $\gamma \! \neq \! \gamma'$
with small action difference ($\mathcal S_\gamma \! -\! \mathcal S_{\gamma'} \!\sim\! \hbar$),
where $\gamma$ forms a loop and $\gamma'$ follows
the loop in opposite direction, while it coincides with $\gamma$
for the rest of the trajectory \cite{Richter2002}. Due to the time-reversed traversal
of the loop the two paths acquire, in the presence of a magnetic field $B$, an additional
action difference $  (\mathcal S_\gamma - \mathcal S_{\gamma'})/\hbar  \!=\!
   4 \pi A B /\Phi_0$, where $A$ is the enclosed (loop) area
 and $\Phi_0$ the flux quantum. Moreover, during the loop $\gamma$ and $\gamma'$
have opposite reflections, $\varphi_j \!=\! -\varphi_j'$,
and hence
\begin{equation}
 K_{\gamma} K^{*}_{\gamma'} = \exp[2 \ci \sum_{j=1}^{n_b} \ \Gamma_{\mathrm{HH}\Uparrow} (\varphi_j)\ ] \, .
\label{KK}
\end{equation}
For  chaotic dynamics in a cavity where the escape length $L_\mathrm{esc}$ is much larger
than the average distance $L_b$ between consecutive bounces we can introduce probability
distributions for the areas $A$ and the phases
$ \sum_{j=1}^{n_b}  \Gamma_{\mathrm{HH}\Uparrow} (\varphi_j)$. Our classical simulations
for both the smooth and the hw case
revealed \cite{Krueckl2010} that the probability distributions of
$ \sum_{j=1}^{n_b}  \Gamma_{\mathrm{HH}\Uparrow} (\varphi_j)$
coincide very well (for $n_b \!>\! 5$ and  $\xi \!<\! 1$) with the distribution
$\tilde{\xi}  \sum_{j=1}^{n_b} \varphi_j$ with a renormalized HH-LH coupling
 $\tilde{\xi}^{\rm Berry} \simeq 0.6 \xi^{\rm Berry}$ and  $\tilde{\xi}^{\rm hw}  \simeq 0.2 \xi^{\rm hw}$.
This allows us to treat both cases on equal footing by replacing Eq.~(\ref{KK}) through
$ K_{\gamma} K^{*}_{\gamma'} = e^{2 \ci\ \tilde{\xi} \alpha}$ with
$\alpha \!=\! \sum^{n_b}_{j=1} \varphi_j $.

\begin{figure}[t]
 \centering \includegraphics[width=\columnwidth]{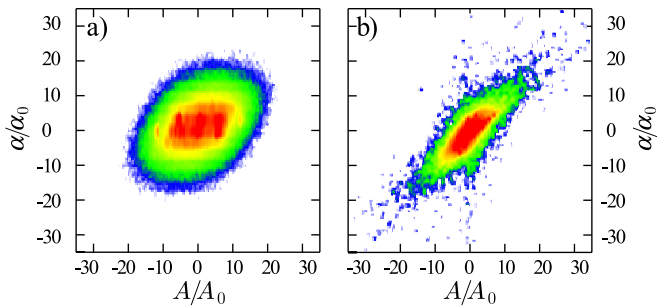}
        \caption{(\emph{Color online})
Probability distributions to find an orbit with
enclosed area $A$ and accumulated angle $\alpha$ for (a) a chaotic cavity
(inset Fig.~\ref{4-geometries}(a)) and (b) a  disc (inset Fig.~\ref{4-geometries}(b)).
(Red (central) regions correspond to high probability). }
\label{P-A-alpha}
\end{figure}

\begin{figure}[t]
   \centering \includegraphics[width=0.8\columnwidth]{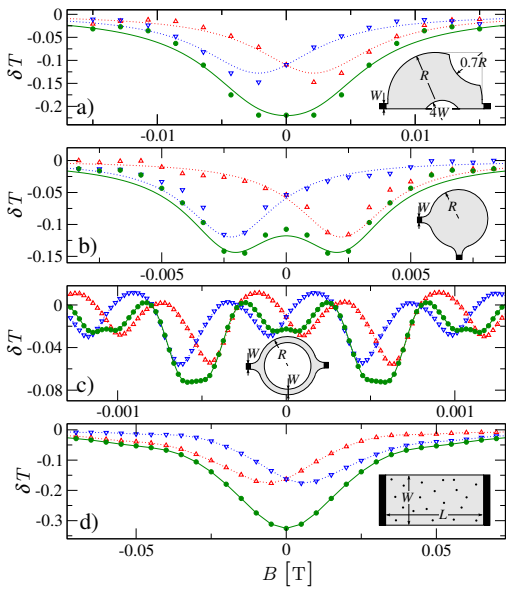}
        \caption{(\emph{Color online})
HH-LH coupling-induced Berry phase effects on weak localization in various mesoscopic
hole gases. The WL correction  $\delta T$ is shown for a ballistic
chaotic cavity (a), disc (b), AB ring (c) and a diffusive strip (d).
Red ($\triangle$) and blue ($\nabla$) triangles denote quantum mechanical
transmissions $\delta T_{\rm U}(B), \delta T_{\rm L}(B)$
adding up to the full $\delta T(B)$ (green bullets, see text below Eq.~(\ref{Landauer})).
The red and blue dotted curves in a) and b) show our semiclassical results
(\ref{lorentzapprox}) with the horizontal displacements given by the Berry field
(\ref{Berryfield}) reflecting geometrical correlations
(based on the calculated classical quantities
$A_0  \simeq 12200 (56129)\mbox{nm}^2$, $\alpha_0 \simeq 1.92 (5.89)$,
 $\rho \! \simeq \! -0.5  (-0.8)$ in a), (b))).
In c), d) the lines are guides to the eye of the quantum results.
Parameters used:
$\gamma_1\!=\!6.85$, $\bar \gamma\!=\!2.5$, $ka/\pi\!=\! 0.64$;
a), b), c): 5 and d) 15 open modes per subblock.
Geometries (lengths in units of nm):
a): $R\!=\!350, W\!=\!40$,
b): $R\!=\!200, W\!=\!40$,
c): $R\!=\!800, W\!=\!40$,
d): $L\!=\!2000, W\!=\!120$.
Averages taken over $\sim2000$  energies and geometries (a-c) and
$\sim 1000$ disorder configurations (d).
}
  \label{4-geometries}
\end{figure}

Generalizing the semiclassical approaches for electron \cite{Baranger1993,Richter2002}
to HH $\Uparrow (\Downarrow)$ transport the WL correction can  then be expressed as an integral over trajectory
lengths,
\begin{equation}
 \delta T_{\rm U(L)} = \frac{\delta T^{(0)} }{L_\mathrm{esc} }
  \int_0^\infty \mathrm{e}^{-L/L_\mathrm{esc}} \mathcal M(L;B,\mp\tilde{\xi})~dL \, .
  \label{lorentzexact}
\end{equation}
Here $\delta T^{(0)}$ is the WL correction for $B\!=\!0, \tilde{\xi} \!=\!0$
($\delta T^{(0)} \!=\! -1/(4\!-\!2/N) $ for a chaotic electronic conductor
\cite{Beenakker1997}), and
\begin{equation}
 \mathcal M(L;B,\tilde{\xi}) \! =\!
   \int_{-\infty}^\infty\! \! \!  dA
   \int_{-\infty}^\infty\!\!\! d\alpha
        P_L(A, \alpha)\\
        e^{2\pi\ci[ \tilde{\xi} \alpha/\pi  +
          2 A B /\Phi_0 ]}  \, ,
\label{M}
\end{equation}
where $ P_L(A, \alpha) $ is the joint probability distribution for the accumulated
areas and angles. While both parameters follow Gaussian distributions, we stress that
there exist non-universal correlations between $A$ and $\alpha$ reflecting the geometry of the quantum dot.
When plotting $ P_L(A, \alpha)$ these correlations show up as deviations from a circular symmetry,
as illustrated in Fig.~\ref{P-A-alpha}(a) showing classical simulations for a chaotic cavity
(inset Fig.~\ref{4-geometries}(a)).

The central limit theorem implies a two-dimensional multivariate normal distribution,
\begin{equation}
P_L (A, \alpha)\! =\! \frac{1}{2 \pi \sigma } \!
 \exp\! \left[ - \frac{ (A/A_0)^2\!+\! (\alpha/\alpha_0)^2\!-\!2 \rho A \alpha / (A_0 \alpha_0)}{2 (1-\rho^2) L/L_b}
 \right]
\label{Aalpha}
\end{equation}
with $\! \sigma\!\!=\!\! A_0 \alpha_0\! \sqrt{(1\!-\! \rho^2)L/L_b} $.\ Correlations are
encoded~in $\rho$ ranging from 0  to $\! \pm 1$.
Assuming ergodicity we obtain
for the variances of the angle $\alpha_0^2\!=\! 4(\pi\!-\!2)$, area
$A_0^2 \! \simeq \! \tfrac{2}{15} [L_b^2 \!+\! {\rm var}(L_b)]^2$
and covariance $\rho A_0 \alpha_0 \!=\! L_b^2(\tfrac{\pi}{4} \!-\! \tfrac{1}{3})$
\cite{Krueckl2010}.
This leads to the geometry-dependent
$\rho \! \simeq \! 0.58 / [1\!+\! {\rm var}(L_b) / L_b^2]$,
{\em i.e.}, $\rho \! < \! 0.58$
for a chaotic system.
($\rho \approx 0.5$ for the cavity in
Fig.~\ref{4-geometries}(a).)
The correlations can be stron\-ger in non-chaotic systems and are
 pronounced for a disk (inset Fig.~\ref{4-geometries}(b))
 as we see in Fig.~\ref{P-A-alpha}(b). (We find $\rho\! \approx\! 0.8$.)

\begin{figure}[t]
 \centering \includegraphics[width=\columnwidth]{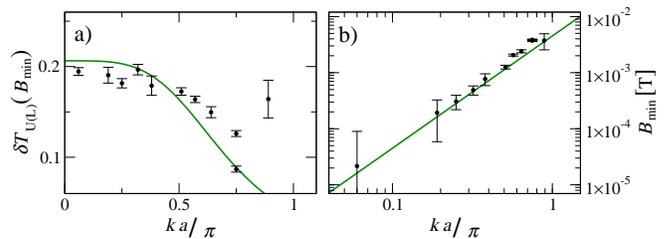}
        \caption{(\emph{Color online})
  Dependence of (a) the depth $\delta T_{U(L)}(B_{\rm min})$ and (b) the position $B_{\rm min}$ of the
  magneto transmission weak localization dip on $ka$ (governing the effective HH-LH coupling,
  see Eq.~(\ref{approxfraction})) for HH transport through a chaotic quantum dot (inset Fig.~\ref{4-geometries}(a)).
Numerical quantum results (symbols)
are compared to the semiclassical predictions (\ref{beta},\ref{Berryfield})
((green) lines)
for $\gamma_1\!=\!6.85$, $\bar \gamma\!=\!2.5$ (for GaAs).
    \label{plotHH}}
\end{figure}

Using Eqs.~(\ref{M},\ref{Aalpha}) we get from Eq.~(\ref{lorentzexact})
semiclassically a  Lorentzian WL dip magneto conductance profile
\begin{equation}
 \delta T_{\rm U(L)} (B) =
   \frac{\beta \delta T^{(0)} }{1 \!+\! [2\pi \sqrt{2\beta} A_0 (B\!\mp\! B_\mathrm{Berry})/\Phi_0)]^2
   L_\mathrm{esc}/L_b}
 \label{lorentzapprox}
\end{equation}
 with a depth $\delta T_{\rm U(L)}(B_{\rm min}) \!=\! \beta \delta T^{(0)}$ with
\begin{equation}
 \beta = [1 + 2 \alpha_0^2 (1- \rho^2) \tilde{\xi}^2 L_\mathrm{esc}/ L_b ]^{-1}  \, .
\label{beta}
\end{equation}
As a main result, the  WL dip is shifted by the Berry field
\begin{equation}
B_\mathrm{Berry} = \rho\ \tilde{\xi} \ \frac{\alpha_0 \Phi_0}{2 \pi A_0} \, ,
 \label{Berryfield}
\end{equation}
which relies on both, {\em quantum} HH-LH coupling $\tilde{\xi}$
and finite {\em classical} $A$-$\alpha$ correlations $\rho$.

In Fig.~\ref{plotHH}(a,b) we compare our predictions (\ref{beta},\ref{Berryfield})
for the dip depth, $\delta T_U(B_{\rm min}) \!=\! -\beta/(4\!-\!2/N) $,
and displacement, $B_{\rm Berry}$, with numerical
recursive Green function calculations \cite{Wimmer2009} of these quantities
for a chaotic quantum dot (inset Fig.~\ref{4-geometries}(a)) for
different HH-LH couplings by tuning the vertical confinement $a$. The quantum results
(symbols) show quantitative agreement with the semiclassical curves (green lines),
which are entirely based on the classical parameters
 $A_0, \alpha_0$ and $\rho$.

Finally, we analyze in the central Fig.~\ref{4-geometries} the effect
of the geometrical correlation $\rho$ on WL in different representa\-tive
mesoscopic systems for fixed, realistic HH-LH coupling.
Panel (a) depicts the WL transmission profile of a chaotic cavity.
Our semiclassical results (without free parameters)
show remarkable agreement with the quantum calculations.
The nonzero $\rho \approx 0.5$ gives rise to a splitting of the
$T_U$ and $T_L$ traces by $2B_{\rm Berry}$ leading to a flattened
WL dip for $T=T_{\rm U} + T_{\rm L}$ compared to the Lorentzian WL profile for electrons.
Panel (b) shows results for the circular dot with larger correlation
($\rho \approx 0.8$). Accordingly, the Berry field
is stronger leading to an WAL-type overall profile.
Correspondingly, we find in the averaged transmission of AB-rings (panel (c))
distinct additional features at $B\!=\!0$ \cite{Zulicke-comment} absent in electron transport.

We close with several remarks:

(i)
Corresponding transport calculations for dots with smooth confinement
yield a $ka$ scaling of $B_{\rm Berry}$ close to the quartic
behavior predicted by $\xi^{\rm Berry}$ from Eq.~(\ref{berryA}).

(ii)
The correlation mechanism is not restricted to ballistic but also relevant in
diffusive systems, as illustrated in Fig.~\ref{4-geometries}(d), leading to
broadening and deviations of the WL profile from that of a digamma function for electrons.

(iii)
If HH and LH states are both occupied and contribute to transport, our
quantum calculations show a {\em vanishing} WL correction both for diffusive
and chaotic ballistic conductors \cite{Krueckl2010} which, as far as we know,
has not been reported before.
Although the full Hamiltonian (\ref{2Dapprox}) obeys TRS
for $B\!=\!0$, transport is governed by the individual subblocks
$\op{\mathcal H}_\mathrm{U}, \op{\mathcal H}_\mathrm{L}$ that
do not possess TRS, and hence WL is suppressed in a 2DHG
with strong coupling between occupied HH and LH states.
It is notable that this kind of effective TRS breaking,
recently discussed in the context of graphene
and topological insulators \cite{Bernevig2006},
is already present in the well-established system of a 2DHG.
Interestingly, if only HH states are occupied,
TRS breaking in each subblock can be traced back to the Berry field
(\ref{Berryfield}), {\em i.e.} system-specific classical corre\-la\-tions
determine the degree of TRS breaking, and hence
the mere knowledge of the overall universality class is insufficient.

(iv)
SOI terms due to SIA and BIA couple the subblocks, eventually restore TRS and give rise to WAL
effects on top of the mechanisms illustrated in Fig.~\ref{4-geometries}; we checked this numerically
for BIA for the diffusive and ballistic case~\cite{Krueckl2010}.  Hence in 2DHG-based AB measurements
such as \cite{Yau2002,Grbic2007} presumably both SOI and HH-LH coupling-induced
phases affect the AB signal. The latter mechanism should be more clearly observable
in systems with reduced SOI such as WL studies in Si \cite{Kuntsevich2007}. Moreover these WAL effects might also be visible in
p-doped ferromagnetic semiconductors such as GaMnAs~\cite{Neu07}. 
From our analysis we expect to observe equivalent WL effects
also in other 2D systems where the band structure gives rise to geometric phases. Promising candidates are
e.g.\ HgTe-based quantum wells with a tunable band topology~\cite{Bernevig2006} that is directly
related to the Berry connection~\cite{Fu2006}.

\begin{acknowledgments}
We acknowledge funding through the Deutsche For\-schungsgemeinschaft
(DFG-JST Forschergruppe on Topological Electronics (KR)
and project KR-2889/2 (VK)), DAAD (MW), TUBA under grant I.A/TUBA-GEBIP/2010-1 (IA) and the A.~v.~Humboldt Foundation (JK).
\end{acknowledgments}

\bibliographystyle{apsrev}

\begin{thebibliography}{31}
\expandafter\ifx\csname natexlab\endcsname\relax\def\natexlab#1{#1}\fi
\expandafter\ifx\csname bibnamefont\endcsname\relax
  \def\bibnamefont#1{#1}\fi
\expandafter\ifx\csname bibfnamefont\endcsname\relax
  \def\bibfnamefont#1{#1}\fi
\expandafter\ifx\csname citenamefont\endcsname\relax
  \def\citenamefont#1{#1}\fi
\expandafter\ifx\csname url\endcsname\relax
  \def\url#1{\texttt{#1}}\fi
\expandafter\ifx\csname urlprefix\endcsname\relax\def\urlprefix{URL }\fi
\providecommand{\bibinfo}[2]{#2}
\providecommand{\eprint}[2][]{\url{#2}}

\bibitem[{\citenamefont{B.~W.~Hapke and Smythe}(1993)}]{Hapke1993}
\bibinfo{author}\bibnamefont{B.~W.~Hapke}, \bibnamefont{R.~M.~Nelson}
  \bibnamefont{and} \bibinfo{author}{\bibfnamefont{W.~D.}
  \bibnamefont{Smythe}}, \bibinfo{journal}{Science}
  \textbf{\bibinfo{volume}{260}}, \bibinfo{pages}{509} (\bibinfo{year}{1993}).

\bibitem[{\citenamefont{van Albada and Lagendijk}(1985)}]{Albada-Wolf}
\bibinfo{author}{\bibfnamefont{M.~P.} \bibnamefont{van Albada}}
  \bibnamefont{and}
  \bibinfo{author}{\bibfnamefont{A.}~\bibnamefont{Lagendijk}},
  \bibinfo{journal}{Phys. Rev. Lett.} \textbf{\bibinfo{volume}{55}},
  \bibinfo{pages}{2692} (\bibinfo{year}{1985});
\bibinfo{author}{\bibfnamefont{P.-E.} \bibnamefont{Wolf}} \bibnamefont{and}
  \bibinfo{author}{\bibfnamefont{G.}~\bibnamefont{Maret}},
 \bibnamefont{{\em ibid,}}
  \bibinfo{pages}{2696} (\bibinfo{year}{1985}).

\bibitem[{\citenamefont{Larose et~al.}(2004)\citenamefont{Larose, Margerin, van
  Tiggelen, and Campillo}}]{Larose2004}
  \bibnamefont{E.~Larose et al.,}
  \bibinfo{journal}{Phys. Rev. Lett.} \textbf{\bibinfo{volume}{93}},
  \bibinfo{pages}{048501} (\bibinfo{year}{2004}).

\bibitem[{\citenamefont{Bayer and Niederdr\"ank}(1993)}]{Bayer1993}
\bibinfo{author}{\bibfnamefont{G.}~\bibnamefont{Bayer}} \bibnamefont{and}
  \bibinfo{author}{\bibfnamefont{T.}~\bibnamefont{Niederdr\"ank}},
  \bibinfo{journal}{Phys. Rev. Lett.} \textbf{\bibinfo{volume}{70}},
  \bibinfo{pages}{3884} (\bibinfo{year}{1993}).

\bibitem[{\citenamefont{Hartung et~al.}(2008)\citenamefont{Hartung, Wellens,
  M\"uller, Richter, and Schlagheck}}]{Hartung2008}
   \bibnamefont{M.~Hartung et al.,}
  \bibinfo{journal}{Phys. Rev. Lett.} \textbf{\bibinfo{volume}{101}},
  \bibinfo{pages}{020603} (\bibinfo{year}{2008}).

\bibitem[{\citenamefont{Abrahams et~al.}(1979)\citenamefont{Abrahams, Anderson,
  Licciardello, and Ramakrishnan}}]{Abrahams1979}
  \bibnamefont{E.~Abrahams et al.,}
  \bibinfo{journal}{Phys. Rev. Lett.}
  \textbf{\bibinfo{volume}{42}}, \bibinfo{pages}{673} (\bibinfo{year}{1979}).

\bibitem[{\citenamefont{Altshuler et~al.,}(1980)\citenamefont{Altshuler,
  Khmel'nitzkii, Larkin, and Lee}}]{Altshuler1980}
  \bibnamefont{B.L.~Altshuler et al.}
  \bibinfo{journal}{Phys. Rev. B} \textbf{\bibinfo{volume}{22}},
  \bibinfo{pages}{5142} (\bibinfo{year}{1980}).

\bibitem[{\citenamefont{S.~Hikami and Nagaoka}(1980)}]{Hikami1980}
\bibinfo{author}{\bibfnamefont{A.~L.} \bibnamefont{S.~Hikami}}
  \bibnamefont{and} \bibinfo{author}{\bibfnamefont{Y.}~\bibnamefont{Nagaoka}},
  \bibinfo{journal}{Prog. Theor. Phys.} \textbf{\bibinfo{volume}{63}},
  \bibinfo{pages}{707} (\bibinfo{year}{1980}).

\bibitem[{\citenamefont{Winkler}(2003)}]{winkler:book}
\bibinfo{author}{\bibfnamefont{R.}~\bibnamefont{Winkler}},
  \emph{\bibinfo{title}{Spin-orbit Coupling Effects in Two-Di\-men\-sional Electron
  and Hole Systems }}
  (\bibinfo{publisher}{Springer},
  \bibinfo{year}{2003}).

\bibitem[{\citenamefont{McPhail et~al.,}(2004)\citenamefont{McPhail, Yasin,
  Hamilton, Simmons, Linfield, Pepper, and Ritchie}}]{McPhail2004}
\bibnamefont{S.~McPhail et al.}
  \bibinfo{journal}{Phys. Rev. B} \textbf{\bibinfo{volume}{70}},
  \bibinfo{pages}{245311} (\bibinfo{year}{2004}).

\bibitem[{\citenamefont{Faniel et~al.}(2007)\citenamefont{Faniel, Hackens,
  Vlad, Moldovan, Gustin, Habib, Melinte, Shayegan, and Bayot}}]{Faniel2007}
\bibnamefont{S.~Faniel et al,,}
  \bibinfo{journal}{Phys. Rev. B} \textbf{\bibinfo{volume}{75}},
  \bibinfo{pages}{193310} (\bibinfo{year}{2007}).

\bibitem[{\citenamefont{Yau et~al.}(2002)\citenamefont{Yau, De~Poortere, and
  Shayegan}}]{Yau2002}
\bibinfo{author}{\bibfnamefont{J.-B.} \bibnamefont{Yau}},
  \bibinfo{author}{\bibfnamefont{E.~P.} \bibnamefont{De~Poortere}},
  \bibnamefont{and} \bibinfo{author}{\bibfnamefont{M.}~\bibnamefont{Shayegan}},
  \bibinfo{journal}{Phys. Rev. Lett.} \textbf{\bibinfo{volume}{88}},
  \bibinfo{pages}{146801} (\bibinfo{year}{2002}).

\bibitem[{\citenamefont{Grbi\ifmmode~\acute{c}\else \'{c}\fi{}
  et~al.}(2007)\citenamefont{Grbi\ifmmode~\acute{c}\else \'{c}\fi{}, Leturcq,
  Ihn, Ensslin, Reuter, and Wieck}}]{Grbic2007}
\bibnamefont{B.~Grbi\ifmmode~\acute{c}\else \'{c}\fi{} et al.,}
  \bibinfo{journal}{Phys. Rev. Lett.} \textbf{\bibinfo{volume}{99}},
  \bibinfo{pages}{176803} (\bibinfo{year}{2007}).

\bibitem[{\citenamefont{J\"a\"askel\"ainen and Z\"ulicke}(2010)}]{Zulicke2010}
\bibinfo{author}{\bibfnamefont{M.}~\bibnamefont{J\"a\"askel\"ainen}}
  \bibnamefont{and}
  \bibinfo{author}{\bibfnamefont{U.}~\bibnamefont{Z\"ulicke}},
  \bibinfo{journal}{Phys. Rev. B} \textbf{\bibinfo{volume}{81}},
  \bibinfo{pages}{155326} (\bibinfo{year}{2010}).

\bibitem[{\citenamefont{Ferry and Goodnick}(1997)}]{Ferry1997}
\bibinfo{author}{\bibfnamefont{D.~K.} \bibnamefont{Ferry}} \bibnamefont{and}
  \bibinfo{author}{\bibfnamefont{S.~M.} \bibnamefont{Goodnick}},
  \emph{\bibinfo{title}{Transport in Nanostructures}}
  (\bibinfo{publisher}{{Cambridge University Press}},
  \bibinfo{address}{Cambridge}, \bibinfo{year}{1997}).

\bibitem[{\citenamefont{Beenakker}(1997)}]{Beenakker1997}
\bibinfo{author}{\bibfnamefont{C.~W.~J.} \bibnamefont{Beenakker}},
  \bibinfo{journal}{Rev. Mod. Phys.} \textbf{\bibinfo{volume}{69}},
  \bibinfo{pages}{731} (\bibinfo{year}{1997}).

\bibitem[{\citenamefont{Luttinger and Kohn}(1955)}]{luttinger1955}
\bibinfo{author}{\bibfnamefont{J.~M.} \bibnamefont{Luttinger}}
  \bibnamefont{and} \bibinfo{author}{\bibfnamefont{W.}~\bibnamefont{Kohn}},
  \bibinfo{journal}{Phys. Rev.} \textbf{\bibinfo{volume}{97}},
  \bibinfo{pages}{869} (\bibinfo{year}{1955}).

\bibitem[{\citenamefont{Berry}(1984)}]{Berry84}
\bibinfo{author}{\bibfnamefont{M.~V.} \bibnamefont{Berry}},
  \bibinfo{journal}{Proc. R. Soc. London A} \textbf{\bibinfo{volume}{392}},
  \bibinfo{pages}{45} (\bibinfo{year}{1984}).

\bibitem[{\citenamefont{Chang and Niu}(1996)}]{Chang96}
\bibinfo{author}{\bibfnamefont{M.-C.} \bibnamefont{Chang}} \bibnamefont{and}
  \bibinfo{author}{\bibfnamefont{Q.}~\bibnamefont{Niu}},
  \bibinfo{journal}{Phys. Rev. B} \textbf{\bibinfo{volume}{53}},
  \bibinfo{pages}{7010} (\bibinfo{year}{1996}).

\bibitem[{\citenamefont{Haldane}(2004)}]{haldane04}
\bibinfo{author}{\bibfnamefont{F.~D.~M.} \bibnamefont{Haldane}},
  \bibinfo{journal}{Phys. Rev. Lett.} \textbf{\bibinfo{volume}{93}},
  \bibinfo{pages}{206602} (\bibinfo{year}{2004}).

\bibitem[{\citenamefont{Chang and Niu}(2008)}]{Chang2008}
\bibinfo{author}{\bibfnamefont{M.-C.} \bibnamefont{Chang}} \bibnamefont{and}
  \bibinfo{author}{\bibfnamefont{Q.}~\bibnamefont{Niu}},
  \bibinfo{journal}{J. Phys.: Cond. Matter}
  \textbf{\bibinfo{volume}{20}}, \bibinfo{pages}{193202}
  (\bibinfo{year}{2008}).

\bibitem[{\citenamefont{Murakami et~al.}(2003)\citenamefont{Murakami, Nagaosa,
  and Zhang}}]{Murakami2003}
\bibinfo{author}{\bibfnamefont{S.}~\bibnamefont{Murakami}},
  \bibinfo{author}{\bibfnamefont{N.}~\bibnamefont{Nagaosa}}, \bibnamefont{and}
  \bibinfo{author}{\bibfnamefont{S.-C.} \bibnamefont{Zhang}},
  \bibinfo{journal}{Science} \textbf{\bibinfo{volume}{301}},
  \bibinfo{pages}{1348} (\bibinfo{year}{2003}).

\bibitem[{\citenamefont{L\"{o}wdin}(1951)}]{loewdin1951}
\bibinfo{author}{\bibfnamefont{P.-O.} \bibnamefont{L\"{o}wdin}},
  \bibinfo{journal}{J. Chem. Phys.} \textbf{\bibinfo{volume}{19}},
  \bibinfo{pages}{1396} (\bibinfo{year}{1951}).

\bibitem[{com()}]{comment-subband}
\emph{\bibinfo{title}{{\rm We checked this 2DHG model by comparing with
  bandstructure calculations based on the full and on a $6\!\times\!
  6$-Hamiltonian including the second transverse HH mode. }}}

\bibitem[{\citenamefont{Broido and Sham}(1985)}]{broido1985}
\bibinfo{author}{\bibfnamefont{D.A.} \bibnamefont{Broido}} \bibnamefont{and}
  \bibinfo{author}{\bibfnamefont{L.J.} \bibnamefont{Sham}},
  \bibinfo{journal}{Phys. Rev. B} \textbf{\bibinfo{volume}{31}},
  \bibinfo{pages}{888} (\bibinfo{year}{1985}).

\bibitem[{\citenamefont{Fisher and Lee}(1981)}]{fischer1981}
\bibinfo{author}{\bibfnamefont{D.~S.} \bibnamefont{Fisher}} \bibnamefont{and}
  \bibinfo{author}{\bibfnamefont{P.~A.} \bibnamefont{Lee}},
  \bibinfo{journal}{Phys. Rev. B} \textbf{\bibinfo{volume}{23}},
  \bibinfo{pages}{6851} (\bibinfo{year}{1981}).

\bibitem[{\citenamefont{Baranger
  et~al.}(1993{\natexlab{a}})\citenamefont{Baranger, Jalabert, and
  Stone}}]{Baranger1993}
\bibinfo{author}{\bibfnamefont{H.~U.} \bibnamefont{Baranger}},
  \bibinfo{author}{\bibfnamefont{R.~A.} \bibnamefont{Jalabert}},
  \bibnamefont{and} \bibinfo{author}{\bibfnamefont{A.~D.} \bibnamefont{Stone}},
  \bibinfo{journal}{Phys. Rev. Lett.} \textbf{\bibinfo{volume}{70}},
  \bibinfo{pages}{3876} (\bibinfo{year}{1993}{\natexlab{a}});
    \bibinfo{journal}{Chaos}
  \textbf{\bibinfo{volume}{3}}, \bibinfo{pages}{665}
  (\bibinfo{year}{1993}{\natexlab{b}}).


\bibitem[{\citenamefont{Richter and Sieber}(2002)}]{Richter2002}
\bibinfo{author}{\bibfnamefont{K.}~\bibnamefont{Richter}} \bibnamefont{and}
  \bibinfo{author}{\bibfnamefont{M.}~\bibnamefont{Sieber}},
  \bibinfo{journal}{Phys. Rev. Lett.} \textbf{\bibinfo{volume}{89}},
  \bibinfo{pages}{206801} (\bibinfo{year}{2002}).

\bibitem{Krueckl2010}
V.~Krueckl et al., unpublished.

\bibitem{Wimmer2009}
M.~Wimmer and K.~Richter. J. Comput. Phys. {\bf 228}, 8548 (2009).

\bibitem{Zulicke-comment}
These are presumably related to features in $h/e$ AB oscillations
recently discussed for a 1d model in Ref.~\cite{Zulicke2010}.

\bibitem{Bernevig2006}
B.~A.~Bernevig, T.~L.~Hughes, and S.~H.~Zhang, Science {\bf 314}, 1757 (2006).

\bibitem{Kuntsevich2007}
A.Yu.~Kuntsevich et al., Phys.~Rev.~B {\bf 75}, 195330 (2007).

\bibitem{Neu07}
D.~Neumaier et al., Phys. Rev. Lett. {\bf 99}, 116803  (2007).

\bibitem{Fu2006}
L.~Fu and C.~L.~Kane. Phys. Rev. B {\bf 74}, 195312 (2006).

\end{thebibliography}

\end{document}